\documentclass[aps,prl,twocolumn,footinbib,bibnotes,preprintnumbers]{revtex4-1} 

\usepackage[utf8]{inputenc}
\usepackage{amsmath,amsfonts,epsfig,color,latexsym}
\usepackage{amssymb}
\usepackage[english, USenglish]{babel}
\usepackage{epsfig,psfrag}
\usepackage{slashed}
\usepackage[normalem]{ulem}

\usepackage{soul,xcolor}
\usepackage{hyperref}
\setstcolor{red}
\usepackage{comment}

\usepackage{subcaption}

\usepackage{stackengine}
\usepackage{tikz}
\usetikzlibrary{positioning,arrows}
\usepgflibrary{arrows.meta}
\usetikzlibrary{decorations.pathmorphing}
\usetikzlibrary{decorations.markings}
\usetikzlibrary{math}
\usetikzlibrary{calc}

\tikzset{
Wilson_1/.style={double distance=1.1pt,postaction={decorate}, decoration={markings,mark=at position .08 with {\arrow{Stealth[scale=0.5]}},mark=at position .985 with {\arrow{Stealth[scale=0.5]}}}},
Scalar/.style={densely dashed},
Gauge/.style={decorate, draw=black, decoration={coil,aspect=0.5, post length = 0pt, pre length = 0pt, segment length=1.75pt,amplitude=1.4pt}},
Gauge-Back/.style={ultra thick, opacity=0.8, decorate, draw=white, decoration={coil,aspect=0.5, post length = 0pt, pre length = 0pt, segment length=1.75pt,amplitude=1.4pt}},
Fermion/.style={},
Fermion-Back/.style={ultra thick, opacity=0.8, draw=white}, 
}

\usepackage{enumitem}

\begin{document}
\newpage
\pagenumbering{arabic}

%%%%%%%%%%%%%%%%%%%%%%%%%%%%%%%%%%%%%

\newcommand{\norm}[1]{{\protect\normalsize{#1}}}
\newcommand{\p}[1]{(\ref{#1})}
\newcommand{\half}{\tfrac{1}{2}}
\newcommand \vev [1] {\langle{#1}\rangle}
\newcommand \ket [1] {|{#1}\rangle}
\newcommand \bra [1] {\langle {#1}|}
\newcommand \pd [1] {\frac{\pa}{\pa {#1}}}
\newcommand \ppd [2] {\frac{\pa^2}{\pa {#1} \pa{#2}}}
\newcommand{\ed}[1]{{\color{red} {#1}}}

\newcommand{\cI}{{\cal I}}
\newcommand{\cM}{{\cal M}} 
\newcommand{\cR}{{\cal R}} 
\newcommand{\cS}{{\cal S}} 
\newcommand{\cK}{{\cal K}}
\newcommand{\cL}{{\cal L}} 
\newcommand{\cF}{{\cal F}}
\newcommand{\cN}{{\cal N}}
\newcommand{\cA}{{\cal A}}
\newcommand{\cB}{{\cal B}}
\newcommand{\cG}{{\cal G}}
\newcommand{\cO}{{\cal O}}
\newcommand{\cY}{{\cal Y}}
\newcommand{\cX}{{\cal X}}
\newcommand{\cT}{{\cal T}}
\newcommand{\cW}{{\cal W}}
\newcommand{\cP}{{\cal P}}
\newcommand{\bP}{{\bar\Phi}}
\newcommand{\mK}{{\mathbb K}}
\newcommand{\nt}{\notag\\} 
\newcommand{\pa}{\partial}
\newcommand{\ep}{\epsilon}
\newcommand{\om}{\omega}
\newcommand{\bom}{\bar\omega}
\newcommand{\etap}{\bar\epsilon}
\newcommand{\vep}{\varepsilon}
\renewcommand{\a}{\alpha}
\renewcommand{\b}{\beta}
\newcommand{\g}{\gamma}
\newcommand{\s}{\sigma}
\newcommand{\la}{\lambda}
\newcommand{\tl}{\tilde\lambda}
\newcommand{\tm}{\tilde\mu}
\newcommand{\tk}{\tilde k}
\newcommand{\da}{{\dot\alpha}}
\newcommand{\db}{{\dot\beta}}
\newcommand{\dg}{{\dot\gamma}}
\newcommand{\dd}{{\dot\delta}}
\newcommand{\q}{\theta}
\newcommand{\bq}{{\bar\theta}}
\renewcommand{\r}{\rho}
\newcommand{\br}{\bar\rho}
\newcommand{\be}{\bar\eta}
\newcommand{\bQ}{\bar Q}
\newcommand{\bx}{\bar \xi}
\newcommand{\tx}{\tilde{x}}
\newcommand{\tr}{\mbox{tr}}
\newcommand{\+}{{\dt+}}
\renewcommand{\-}{{\dt-}}
\newcommand{\ti}{{\textup{i}}}

\newcommand{\dlog}{d{\rm log}}
\newcommand{\tred}[1]{\textcolor{red}{\bfseries #1}}
\newcommand{\eps}{\epsilon}

\newcommand{\api}{{\left(\frac{\alpha}{\pi}\right)}}

\preprint{MPP-2020-109, P3H-20-033, SI-HEP-2020-16}

\title{
The full angle-dependence of the  four-loop cusp anomalous dimension in QED
}

\author{R.\ Br\"{u}ser}
\email{robin.brueser@uni-siegen.de}
\affiliation{Theoretische Physik 1,
 Naturwissenschaftlich-Technische Fakult\"at, 
Universit\"at Siegen, 
57068 Siegen, Germany}

\author{C.\ Dlapa}
\email{dlapa@mpp.mpg.de}
\affiliation{Max-Planck-Institut f{\"u}r Physik, Werner-Heisenberg-Institut, 80805 M{\"u}nchen, Germany}

\author{J.\ M.\ Henn}
\email{henn@mpp.mpg.de}
\affiliation{Max-Planck-Institut f{\"u}r Physik, Werner-Heisenberg-Institut, 80805 M{\"u}nchen, Germany}

\author{K.\ Yan}
\email{kyan@mpp.mpg.de}
\affiliation{Max-Planck-Institut f{\"u}r Physik, Werner-Heisenberg-Institut, 80805 M{\"u}nchen, Germany}

\begin{abstract}
The angle-dependent cusp anomalous dimension governs divergences coming from soft gluon exchanges between heavy particles, such as top quarks. We focus on the matter-dependent contributions and compute the first truly non-planar terms. They appear at four loops and are proportional to a quartic Casimir operator in color space. Specializing our general gauge theory result to U(1), we obtain the full QED four-loop angle-dependent cusp anomalous dimension. While more complicated functions appear at intermediate steps, the analytic answer depends only on multiple polylogarithms with singularities at fourth roots of unity. It can be written in terms of four rational structures, and contains functions of up to maximal transcendental weight seven. Despite this complexity, we find that numerically the answer is tantalizingly close to the appropriately rescaled one-loop formula, 
over most of the kinematic range. We take several limits of our analytic result, which serves as a check and allows us to obtain new, power-suppressed terms. In the anti-parallel lines limit, which corresponds to production of two massive particles at threshold, we find 
that the subleading power correction vanishes.
Finally, we compute the quartic Casimir contribution for scalars in the loop. Taking into account a supersymmetric decomposition, we derive the first non-planar corrections to the quark anti-quark potential in maximally supersymmetric gauge theory.
\end{abstract}

\maketitle

\section{Introduction}

Low-energy gluons lead to large contributions in gauge theory scattering amplitudes and cross sections. 
The ability to predict and resum these enhanced terms is critical for phenomenology. 
It has been understood that scattering amplitudes factorize into a process-dependent finite part and a universal divergent part \cite{Weinberg:1965nx, *STERMAN1987310, *Collins:1989gx}. 
The latter is determined by a set of anomalous dimensions and by the soft anomalous dimension matrix, which describes the renormalization of products of Wilson lines pointing in the particles' directions \cite{Korchemsky:1991zp, *Dixon:2008gr}.
A deeper understanding of infrared divergences is critical for developing manifestly finite approaches to cross sections \cite{ArkaniHamed:2010gh, *Gnendiger:2017pys, *Anastasiou:2018rib, *Hannesdottir:2019rqq}, 
or for deriving quantitative predictions for the finite part of scattering amplitudes from symmetries of the S-matrix \cite{Drummond:2007au, *Chicherin:2018ubl}.

Many studies have focused on the planar limit of massless gauge theories, where significant simplifications occur \cite{Bern:2005iz,Almelid:2015jia}, and where one may glean insights from integrability \cite{Beisert:2006ez}. 
However, non-planar terms in QCD may be numerically dominant, and finite mass effects are important in many situations, for example when producing a top quark pair near threshold. 

The multi-parton soft anomalous dimension matrix is known to two loops \cite{Mitov:2009sv, *Ferroglia:2009ep, *Chien:2011wz} and exhibits a simple dependence on the scattering angles. Whether or not this continues at higher loops is an open question. We will provide further insights into this issue by exploring the two-line case, which is given by the cusp anomalous dimension. 

The cusp anomalous dimension depends on the Euclidean angle $\phi$ formed by the two Wilson lines. It is known to three loops \cite{Korchemsky:1987wg,Grozin:2014hna,Grozin:2015kna}. Thanks to non-Abelian exponentiation, it depends on a reduced set of color factors, and truly non-planar corrections appear for the first time at four loops. 
% Internal note: the nf CF CA^2 term at four loops also contains non-planar Feynman diagrams; however, it can be obtained from all the planar diagrams, plus the knowledge of the quartic Casimir contributions.
They multiply quartic Casimir operators, and have recently been computed for the case of massless particles \cite{Lee:2019zop, Henn:2019rmi, Henn:2019swt, Huber:2019fxe, vonManteuffel:2020vjv}, thereby settling earlier questions about Casimir scaling. In this Letter, we consider the massive case and determine the matter-dependent quartic Casimir term in QCD. This determines the complete four-loop cusp anomalous dimension in QED.

Our work sheds new light on an observed iterative pattern, which allows to predict matter-dependent terms from lower loop orders \cite{Grozin:2014hna,Grozin:2015kna}.
Previous work showed that at four loops this pattern does not hold for all color structures \cite{Grozin:2017css}. The term we computed is precisely the one that captures the deviations from the conjecture in QED. We compare it quantitatively to the conjectured one, beyond the previously known limits.

We also derive new results in several physically important limits. 
Interestingly, we find that the first power-suppressed terms in the 
quark-antiquark threshold limit vanish. 
Moreover, we determine the first non-planar correction to the three-loop quark anti-quark potential in $\mathcal{N}=4$ super Yang-Mills (sYM). 

\section{Methodology}

We compute the vacuum expectation value of the off-shell Wilson lines in momentum space, using the conventions of heavy quark effective theory (HQET) \cite{Grozin:2015kna}. We consider the  gauge group $SU(N_{c})$, see \cite{Bruser:2019auj} for our conventions, and perform calculations in covariant gauge, which allows us to check gauge invariance of the results. We obtain results for QED by adjusting the QCD color factors according to $C_{A} =  d_{R} d_{A}/N_{R} = 0$ and $C_{F}=T_{F}=d_{R} d_{F}/N_{R}=1$ \cite{vanRitbergen:1998pn, *Herzog:2017ohr}, and by replacing the strong coupling strength $\alpha_s = g_{YM}^2/(4\pi)$ by the fine structure constant $\alpha$.

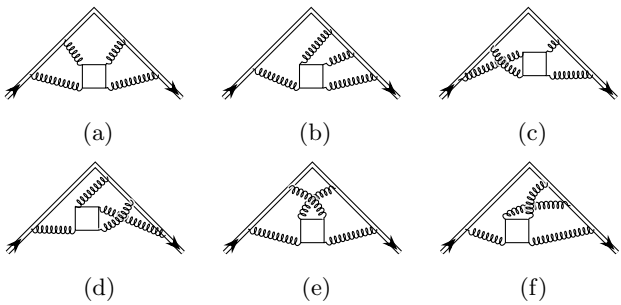
\begin{figure}[t]
\centering
\begin{subfigure}{0.155\textwidth}
 {\begin{tikzpicture}[scale=0.55, baseline=(current bounding box.center)]
% >>>>>>>>>>>>>Topo nr 2 <<<<<<<<<<<

      \node at (0,0) {~};
      \node at (0,-2.1) {~};
	  \tikzmath{\cx=0; \cy=-1.6; \crad =0.4; };
      \coordinate (WL) at (-135:3)   {};
      \coordinate (cusp) at (0,0)   {};
      \coordinate (WR) at (-45:3) {};
      
      \coordinate (B1) at ($(\cx,\cy)+(-135:\crad)$) {};
      \coordinate (B2) at ($(\cx,\cy)+(+135:\crad)$) {};
      \coordinate (B3) at ($(\cx,\cy)+(+45:\crad)$) {};
      \coordinate (B4) at ($(\cx,\cy)+(-45:\crad)$) {};
      
      \coordinate (W1) at (-135:2.2) {};
      \coordinate (W2) at (-135:1) {};
      \coordinate (W3) at (-45:1) {};
      \coordinate (W4) at (-45:2.2) {};
      
      \draw[Wilson_1] (WL)  -- (cusp) -- (WR);
      \draw[Fermion] (B1) -- (B2) -- (B3) -- (B4) -- (B1);
      \draw[Gauge] (B1) -- (W1) ;
      \draw[Gauge] (B2) -- (W2) ;
      \draw[Gauge] (W3) -- (B3) ;
      \draw[Gauge] (W4) -- (B4) ;
\end{tikzpicture}}
 \caption{}
\end{subfigure} 
\begin{subfigure}{0.155\textwidth}
 {\begin{tikzpicture}[scale=0.55, baseline=(current bounding box.center)]
% >>>>>>>>>>>>> Topo nr 3 <<<<<<<<<<<

      \node at (0,0) {~};
      \node at (0,-2.1) {~};
	  \tikzmath{\cx=0; \cy=-1.6; \crad =0.4; };
      \coordinate (WL) at (-135:3)   {};
      \coordinate (cusp) at (0,0)   {};
      \coordinate (WR) at (-45:3) {};
      
      \coordinate (B1) at ($(\cx,\cy)+(-135:\crad)$) {};
      \coordinate (B2) at ($(\cx,\cy)+(+135:\crad)$) {};
      \coordinate (B3) at ($(\cx,\cy)+(+45:\crad)$) {};
      \coordinate (B4) at ($(\cx,\cy)+(-45:\crad)$) {};
      
      \coordinate (W1) at (-135:2) {};
      \coordinate (W2) at (-45:0.65) {};
      \coordinate (W3) at (-45:1.45) {};
      \coordinate (W4) at (-45:2.25) {};
      
      \draw[Wilson_1] (WL)  -- (cusp) -- (WR);
      \draw[Fermion] (B1) -- (B2) -- (B3) -- (B4) -- (B1);
      \draw[Gauge] (B1) -- (W1) ;
      \draw[Gauge] (W2) -- (B2) ;
      \draw[Gauge] (W3) -- (B3) ;
      \draw[Gauge] (W4) -- (B4) ;
\end{tikzpicture}}
  \caption{}
\end{subfigure} 
\begin{subfigure}{0.155\textwidth}
 {\begin{tikzpicture}[scale=0.55, baseline=(current bounding box.center)]
% >>>>>>>>>>>>> Topo nr 4 <<<<<<<<<<<

      \node at (0,0) {~};
      \node at (0,-2.1) {~};
	  \tikzmath{\cx=0.15; \cy=-1.3; \crad =0.4; };
      \coordinate (WL) at (-135:3)   {};
      \coordinate (cusp) at (0,0)   {};
      \coordinate (WR) at (-45:3) {};
      
      \coordinate (B1) at ($(\cx,\cy)+(-135:\crad)$) {};
      \coordinate (B2) at ($(\cx,\cy)+(+135:\crad)$) {};
      \coordinate (B3) at ($(\cx,\cy)+(+45:\crad)$) {};
      \coordinate (B4) at ($(\cx,\cy)+(-45:\crad)$) {};
      
      \coordinate (W1) at (-135:2.4) {};
      \coordinate (W2) at (-135:1.2) {};
      \coordinate (W3) at (-45:1) {};
      \coordinate (W4) at (-45:2.2) {};
      
      \draw[Wilson_1] (WL)  -- (cusp) -- (WR);
      \draw[Fermion] (B1) -- (B2) -- (B3) -- (B4) -- (B1);

      \draw[Gauge] (B2) -- (W1) ;
      \draw[Gauge-Back] (B1) .. controls (-0.5,-1.4) and (-0.7,-1.3) .. (W2) ;
      \draw[Gauge] (B1) .. controls (-0.5,-1.4) and (-0.7,-1.3) .. (W2) ;
      \draw[Gauge] (W3) -- (B3) ;
      \draw[Gauge] (W4) -- (B4) ;
\end{tikzpicture}}
  \caption{}
\end{subfigure} 
\newline
\begin{subfigure}{0.155\textwidth}
 {\begin{tikzpicture}[scale=0.55, baseline=(current bounding box.center)]
% >>>>>>>>>>>>> Topo nr 5 <<<<<<<<<<<

      \node at (0,0) {~};
      \node at (0,-2.1) {~};
	  \tikzmath{\cx=-0.2; \cy=-1.3; \crad =0.4; };
      \coordinate (WL) at (-135:3)   {};
      \coordinate (cusp) at (0,0)   {};
      \coordinate (WR) at (-45:3) {};
      
      \coordinate (B1) at ($(\cx,\cy)+(-135:\crad)$) {};
      \coordinate (B2) at ($(\cx,\cy)+(+135:\crad)$) {};
      \coordinate (B3) at ($(\cx,\cy)+(+45:\crad)$) {};
      \coordinate (B4) at ($(\cx,\cy)+(-45:\crad)$) {};
      
      \coordinate (W1) at (-135:2.2) {};
      \coordinate (W2) at (-45:0.4) {};
      \coordinate (W3) at (-45:1.2) {};
      \coordinate (W4) at (-45:2.4) {};

      \draw[Wilson_1] (WL)  -- (cusp) -- (WR);
      \draw[Fermion] (B1) -- (B2) -- (B3) -- (B4) -- (B1);
      \draw[Gauge] (B1) -- (W1) ;
      \draw[Gauge] (W2) -- (B2) ;
      \draw[Gauge] (W4) -- (B3) ;
      \draw[Gauge-Back] (W3) .. controls (+0.7,-1.3) and (+0.5,-1.4) .. (B4) ;
      \draw[Gauge] (W3) .. controls (+0.7,-1.3) and (+0.5,-1.4) .. (B4) ;

\end{tikzpicture}}
 \caption{}
\end{subfigure} 
\begin{subfigure}{0.155\textwidth}
 {\begin{tikzpicture}[scale=0.55, baseline=(current bounding box.center)]
% >>>>>>>>>>>>> Topo nr 6 <<<<<<<<<<<

      \node at (0,0) {~};
      \node at (0,-2.1) {~};
	  \tikzmath{\cx=0; \cy=-1.6; \crad =0.4; };
      \coordinate (WL) at (-135:3)   {};
      \coordinate (cusp) at (0,0)   {};
      \coordinate (WR) at (-45:3) {};
      
      \coordinate (B1) at ($(\cx,\cy)+(-135:\crad)$) {};
      \coordinate (B2) at ($(\cx,\cy)+(+135:\crad)$) {};
      \coordinate (B3) at ($(\cx,\cy)+(+45:\crad)$) {};
      \coordinate (B4) at ($(\cx,\cy)+(-45:\crad)$) {};
      
      \coordinate (W1) at (-135:2.2) {};
      \coordinate (W2) at (-135:0.8) {};
      \coordinate (W3) at (-45:0.8) {};
      \coordinate (W4) at (-45:2.2) {};
      
      \draw[Wilson_1] (WL)  -- (cusp) -- (WR);
      \draw[Fermion] (B1) -- (B2) -- (B3) -- (B4) -- (B1);
      \draw[Gauge] (B1) -- (W1) ;
      \draw[Gauge] (B2) .. controls (-0.1,-0.8) and (0.2,-0.7) .. (W3) ;
      \draw[Gauge-Back] (W2) .. controls (-0.2,-0.7) and (+0.1,-0.8) .. (B3) ;
      \draw[Gauge] (W2) .. controls (-0.2,-0.7) and (+0.1,-0.8) .. (B3) ;
      \draw[Gauge] (W4) -- (B4) ;
\end{tikzpicture}}
  \caption{}
\end{subfigure} 
\begin{subfigure}{0.155\textwidth}
 {\begin{tikzpicture}[scale=0.55, baseline=(current bounding box.center)]
% >>>>>>>>>>>>> Topo nr 7 <<<<<<<<<<<

      \node at (0,0) {~};
      \node at (0,-2.1) {~};
	  \tikzmath{\cx=-0.3; \cy=-1.6; \crad =0.4; };
      \coordinate (WL) at (-135:3)   {};
      \coordinate (cusp) at (0,0)   {};
      \coordinate (WR) at (-45:3) {};
      
      \coordinate (B1) at ($(\cx,\cy)+(-135:\crad)$) {};
      \coordinate (B2) at ($(\cx,\cy)+(+135:\crad)$) {};
      \coordinate (B3) at ($(\cx,\cy)+(+45:\crad)$) {};
      \coordinate (B4) at ($(\cx,\cy)+(-45:\crad)$) {};
      
      \coordinate (W1) at (-135:2.2) {};
      \coordinate (W2) at (-45:0.6) {};
      \coordinate (W3) at (-45:1.4) {};
      \coordinate (W4) at (-45:2.2) {};

      \draw[Wilson_1] (WL)  -- (cusp) -- (WR);
      \draw[Fermion] (B1) -- (B2) -- (B3) -- (B4) -- (B1);
      \draw[Gauge] (B1) -- (W1) ;
      \draw[Gauge] (B2) .. controls (-0.5,-1.1) and (0.2,-0.9) .. (W3) ;
      \draw[Gauge-Back] (W2) .. controls (-0.1,-0.7) and (+0.05,-1) .. (B3) ;
      \draw[Gauge] (W2) .. controls (-0.1,-0.7) and (+0.05,-1) .. (B3) ;
      \draw[Gauge] (W4) -- (B4) ;
\end{tikzpicture}}
  \caption{}
\end{subfigure} 
   \caption{{Feynman diagrams contributing to the four-loop quartic Casimir term $B$ in eq.~(\ref{definitionB}).}}
   \label{fig:qedquartic}
\end{figure}

We are interested in the matter-dependent quartic Casimir term of the four-loop cusp anomalous dimension,
\begin{align}\label{definitionB}
\Gamma_{\rm cusp}|_{\alpha_{s}^4} =  \left(\frac{\alpha_{s}}{\pi}\right)^4 \frac{d_{R} d_{F}}{N_{R}}  \left[ n_{f} B(x) + n_{s} C(x) \right] + \ldots \,,
\end{align}
where $B(x)$ and $C(x)$ are the functions we wish to determine, $x=e^{i\phi}$, and where the number of (light) fermions and scalars (canonically coupled to the gauge field) is denoted by $n_{f}$ and $n_{s}$, respectively. The dots stand for other color structures \cite{Bruser:2019auj} which we do not consider.
The Feynman diagrams relevant for the calculation of $B(x)$ are shown in Fig.~\ref{fig:qedquartic}. 
We evaluate them, and the ones needed for $C(x)$, in dimensional regularization, with $D=4-2 \eps$. The cusp anomalous dimension is obtained from the coefficient of the $1/\eps$ pole  \cite{Bruser:2019auj}.

We relate the integrals appearing in the Feynman diagrams to a conveniently chosen integral basis \cite{Henn:2013pwa}, and then determine the basis integrals by the method of differential equations (DEs). In the process, we need to handle large systems of linear equations representing integration by parts identities, which we generate and solve using the codes FIRE6 \cite{Smirnov:2019qkx} and LiteRed \cite{Lee:2013mka}. For most of the calculations we use FIRE6's finite field methods, and reconstruct the full $x$ and $D$ dependence using the techniques of \cite{Peraro:2016wsq}. We reconstruct only the DEs and the Feynman diagrams we need, as opposed to creating integral tables. Thanks to choosing an improved basis, relatively few finite field evaluations are required for this step.

The DEs for the integral families shown in Fig.~\ref{fig:qedquartic} 
 involve of the order of $500$ integrals (needed to describe all integrals sharing the same or fewer propagators), with coupled sub-systems of size up to $17$, and denominators of up to degree-20 polynomials in $x$ and $D$. 
We wish to find an integral basis that significantly simplifies the differential equations \cite{Henn:2013pwa}. To solve this complicated problem in an automated way we develop further the algorithm of \cite{Dlapa:2020cwj} that relies on only a partial knowledge of a canonical basis. 

To provide input for the algorithm, we use several ideas: for up to nine propagators, we find candidate uniform weight HQET integrals by a position-space analysis \cite{Grozin:2015kna}, while for integrals with ten or more propagators we perform a leading singularity analysis with the help of the algorithm \cite{Henn:2019swt}. Moreover, we adapt \cite{Dlapa:2020cwj} to scan over a larger set of candidate integrals automatically, and to use knowledge of all basis integrals that are known to be uniform weight. Finally, we also make use of the complementary method \cite{Lee:2014ioa,Gituliar:2017vzm}.
Given the information found by this means, we first transform the diagonal blocks (this is equivalent to considering integrals where propagators are replaced by delta functions) of the DE into canonical form. In a second step, we transform the off-diagonal blocks of the full DEs. 

In this way we obtain a canonical form of the DEs for family (a), (b), (d) and (f),
\begin{align}\label{canonicalDE}
d \vec{f}(x,\eps) = \eps \sum\nolimits_{k}  {\bf m}_{k}  \left[ d\log \alpha_{k}(x) \right] \vec{f}(x,\eps)\,,
\end{align}
where $\vec{f}$ is the vector of basis integrals for each integral family, ${\bf m}_{k}$ are matrices with constant entries, and 
$\vec{\alpha} = \{ x, 1+x,1-x, 1+x^2 , 1-x+x^2,  
\frac{1- \sqrt{-x}}{1+\sqrt{-x}} , \frac{1- \sqrt{-x}+x}{1+\sqrt{-x} +x} 
 \}.$
Setting $x=-z^2$, we solve eq. (\ref{canonicalDE}) in terms of multiple polylogarithms \cite{Duhr:2019wtr} to the order in $\eps$ needed. The boundary values are taken from \cite{Grozin:2017css}, or are determined from physical consistency conditions.

\begin{figure}[t]
 \begin{tikzpicture}[scale=0.6, baseline=(current bounding box.center)]
% >>>>>>>>>>>>>Topo nr 2 <<<<<<<<<<<

      \node at (0,0) {~};
      \node at (0,-2.1) {~};
      \coordinate (WL) at (-135:3)   {};
      \coordinate (cusp) at (0,0)   {};
      \coordinate (WR) at (-45:3) {};      
      
      \coordinate (W1) at (-135:2.2) {};
      \coordinate (W2) at (-135:1) {};
      \coordinate (W3) at (-45:1) {};
      \coordinate (W4) at (-45:2.2) {};
      
      \coordinate (B1) at ($(W2)+(0,-1.2)$) {};
      \coordinate (B2) at ($(W3)+(0,-1.2)$) {};
      
      \draw[Wilson_1] (WL)  -- (cusp) -- (WR);
      \draw[Fermion] (W1) -- (B1) -- (B2) -- (W4);
      \draw[Fermion] (W2) -- (B2);
      \draw[Fermion] (B1) -- (W2);    
 %     \draw[Fermion-Back] (B1) -- (W3);
      \filldraw[white] (0,-1.3) circle (0.08);
      \draw[Fermion] (B1) -- (W3);

\end{tikzpicture}
     \caption{Integral sector involving special functions that do not appear in the final result.}
        \label{fig:sector15}
\end{figure}
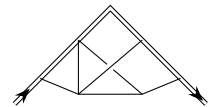
 
For the families (c) and (e), we proceed in the same way for all integrals up to nine propagators, and for all diagonal blocks, except for the ten-propagator sector shown in Fig.~\ref{fig:sector15}.
For this sector we suspect that algebraic basis transformations are insufficient to bring them into canonical form (see Appendix). A canonical form can likely be achieved by enlarging the function space \cite{Duhr:2019wtr}, but for the scope of this paper we can proceed in a simpler way. 

We find that the issue can be bypassed in the computation of the cusp anomalous dimension, which only sees the $1/\epsilon$ pole of each HQET diagram. 
This suggests that complicated terms in individual basis integrals might drop out of the final answer.
Having this in mind, we organize the basis in a way such that each basis element only needs to be computed to $O(\epsilon^0)$. Remarkably,  the new basis forms a closed differential system up to $O(\epsilon)$ corrections  (which are irrelevant for $\Gamma_{\rm cusp}$), $ d \vec{f} - d\, \mathbf{a} (\epsilon, x)\, \vec{f}  =  O(\epsilon)$, where the matrix ${\mathbf a}$ is finite as $\eps \to 0$, and it is nilpotent. 
To the order needed, the solution is given by multiple polylogarithms.

% Plots

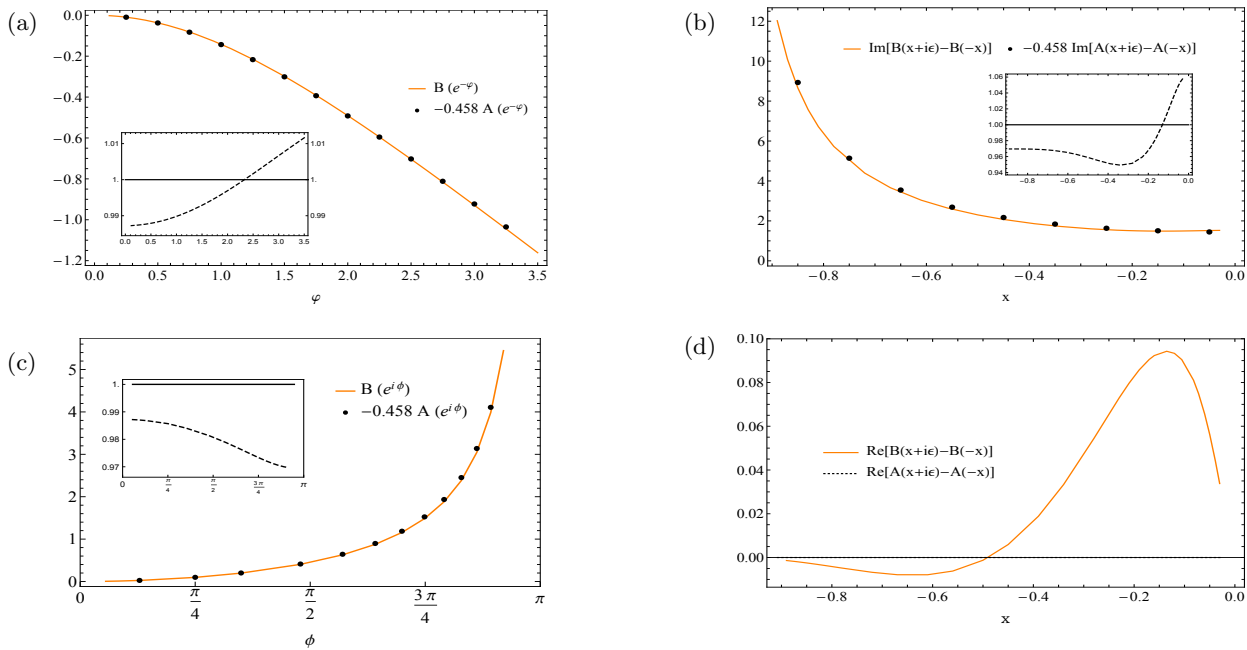
\begin{figure*}[t]
  \begin{minipage}[t]{1.125\textwidth}
  \end{minipage}
  \begin{minipage}[t]{0.49\textwidth}
  \begin{flushleft}
  \begin{tikzpicture}
  \node[inner sep=0pt] at (0,0) {\includegraphics[width=0.75\textwidth,height=0.45\textwidth]{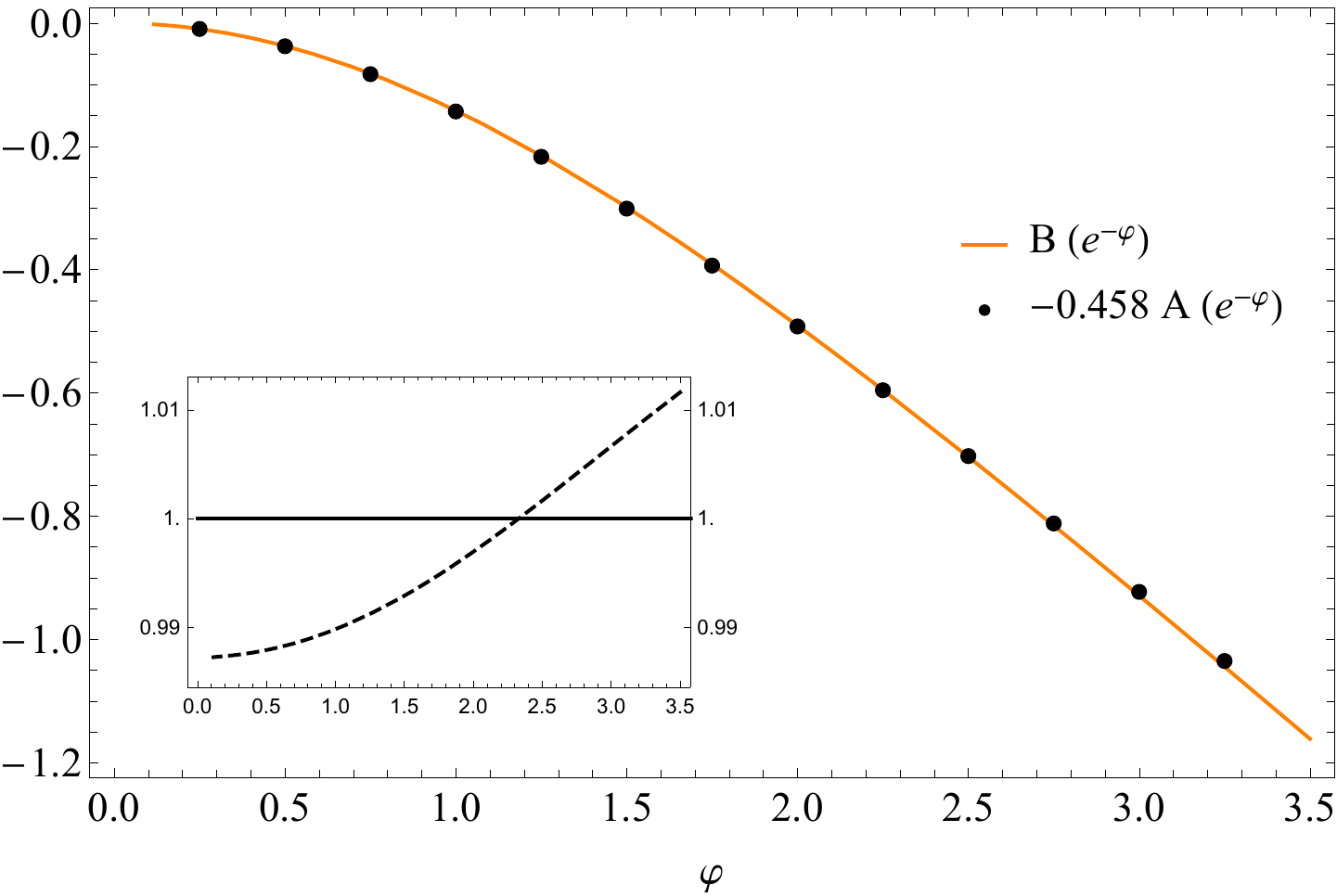}};
  \node at (-0.42\textwidth,0.2\textwidth) {(a)};
  \end{tikzpicture} \\
  \vspace*{3ex}
  \begin{tikzpicture}
  \node[inner sep=0pt] at (0.1,0) {\includegraphics[width=0.87\textwidth,height=0.5\textwidth]{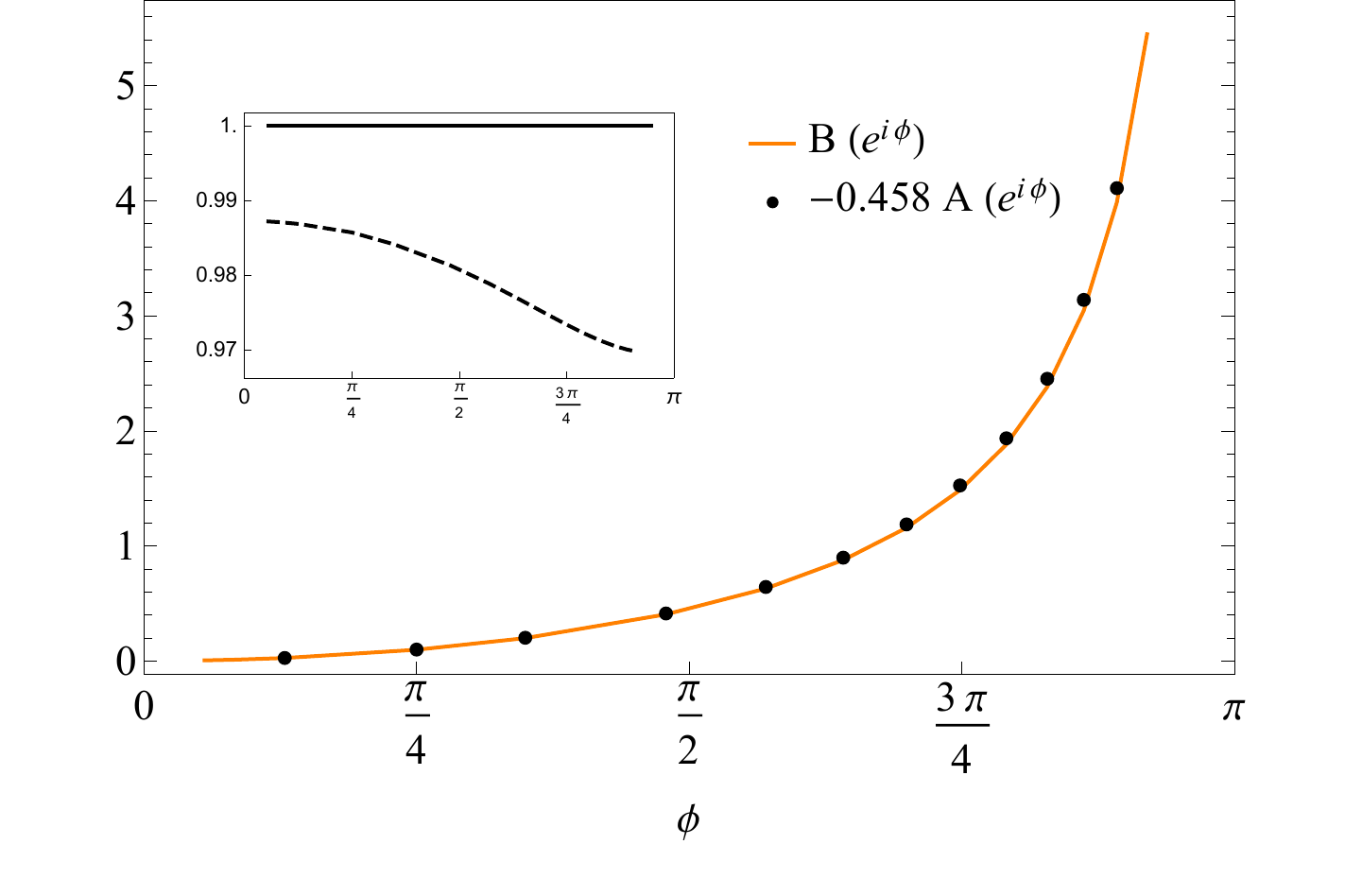}};
  \node at (-0.42\textwidth,0.21\textwidth) {(c)};
  \end{tikzpicture}
  \end{flushleft}
  \end{minipage}
  \begin{minipage}[t]{1.125\textwidth}
  \end{minipage}
  \begin{minipage}[t]{0.49\textwidth}
  \begin{flushleft}
  \begin{tikzpicture}
  \node[inner sep=0pt] at (0,0) {\includegraphics[width=0.75\textwidth,height=0.45\textwidth]{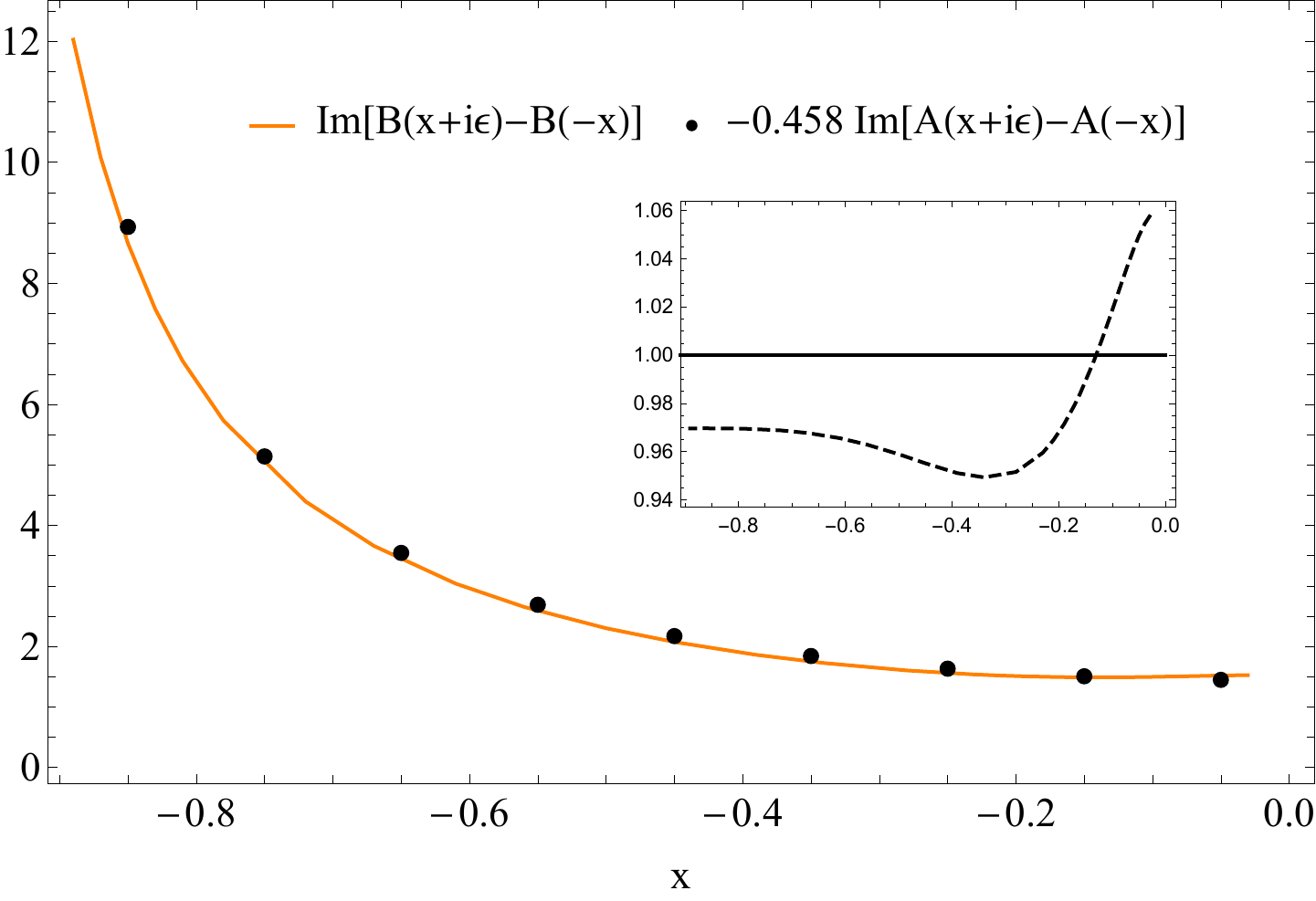}};
  \node at (-0.45\textwidth,0.2\textwidth) {(b)};
  \end{tikzpicture} \\
  \vspace*{2ex}
  \begin{tikzpicture}
  \node[inner sep=0pt] at (0,0) {\includegraphics[width=0.77\textwidth,height=0.45\textwidth]{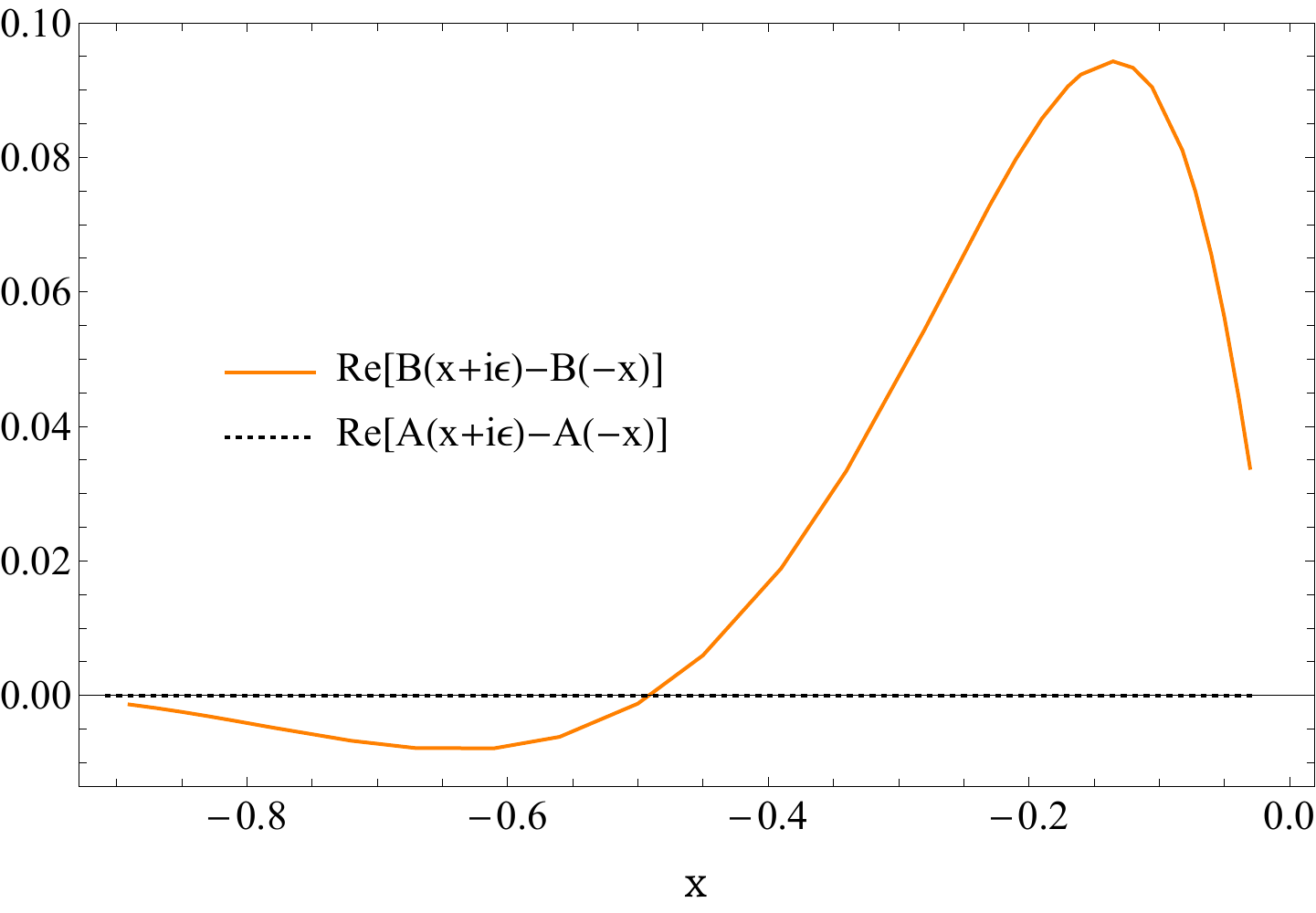}};
  \node at (-0.44\textwidth,0.2\textwidth) {(d)};
  \end{tikzpicture}
  \end{flushleft}
  \end{minipage}
  \caption{{The shape of the four-loop term $B$ agrees well with the rescaled one-loop function $A$. The dashed line in the insets shows their ratio, which agrees with unity within a few per cent, except for (d). 
The plots show different kinematic regions, for Minkowskian angle $\varphi = -i \phi$ (a), Euclidean angle $\phi$ (c), and for $x\in (-1,0)$ (b,d). 
  }}
        \label{fig:plots}
   \end{figure*}

\section{Results}

\subsection{Full result for the QED cusp anomalous dimension}

We obtain the following result for the four-loop cusp anomalous dimension in QED,
\begin{align}\label{cuspQED}
\Gamma_{\rm cusp}(x,\alpha)  = \gamma(\alpha) A(x) + \left(\frac{\alpha}{\pi}\right)^4 n_{f} 
B(x) + \mathcal{O}(\alpha^5) \,,
\end{align}
where 
\begin{align}
\gamma(\alpha) =& \, \api - \frac{5 n_f}{9} \api^2 +  \left(- \frac{ n_f^2}{27} - \frac{55 n_f}{48}  +  n_f \zeta_3  \right)  \api^3\nonumber \\
&+ \left\lbrack  n_f^3 \left( -\frac{1}{81} +\frac{2 \zeta_3}{27} \right) +  n_f^2 \left( \frac{299}{648} +\frac{ \pi^4}{180} - \frac{10 \zeta_3}{9} \right)   \right.
\nonumber \\   & \left. \quad +   n_f \left(  \frac{143}{288} + \frac{37 \zeta_3}{24} - \frac{5 \zeta_5}{2}  \right) \right\rbrack \api^4\,,
\end{align}
and
\begin{align}\label{functionA}
A= -\frac{1+x^2}{1-x^2} \log x  - 1\,,
\end{align}
 is the one-loop function. 
 
The first term in eq. (\ref{cuspQED}) 
comes from propagator-type diagrams \cite{Gracey:1994nn, *Beneke:1995pq},
 while the term $B$, our main new result, comes from the Feynman diagrams shown in Fig.~\ref{fig:qedquartic}.
We write it in the following way
\begin{align}\label{functionB}
B =& \frac{1+x^2}{1-x^2} B_{1} + \frac{x}{1-x^2} B_{2}+ \frac{1-x^2}{x} B_{3} + B_{4}\,,
\end{align}
that makes the rational dependence on $x$ manifest.
The functions $B_{i}$ are given by linear combinations (in $\mathbb{Q}$) of multiple polylogarithms of transcendental weight four to seven.
We provide them as computer-readable ancillary files.
Remarkably, only four of the integration kernels of eq. (\ref{canonicalDE}) appear, namely $\alpha = \{x, 1\pm x, 1+x^2 \}$.

Previously it was conjectured \cite{Grozin:2014hna} that $B(x)$ equals 
\begin{align}\label{conjectureBA}
B_{\rm c}(x) = \left( \frac{\pi^2}{6} - \frac{\zeta_3}{3} - \frac{5 \zeta_5}{3} \right)  A(x) \approx -0.484 \,A(x)\,.
\end{align}
This was found to be inconsistent with the small angle expansion, but curiously it approximately agrees numerically with the exact answer \cite{Grozin:2017css}. 
We can now evaluate our exact result $B(x) - B_{c}(x)$ for any value of $x$, using \cite{Bauer:2000cp}.
Remarkably, we find that the shape of $B(x)$ is very well described by $B_{c}(x)$.
It turns out that one can 
improve the quantitative agreement by adjusting the proportionality constant in eq. (\ref{conjectureBA}).
Fig.~\ref{fig:plots}(a) and Fig.~\ref{fig:plots}(c) show that $-0.458 A(x)$ approximates $B(x)$ within $6 \%$ in a large part of the kinematic regions for
Minkowskian angles $\varphi = -i \phi$, and Euclidean angles, respectively, and similarly for the imaginary part in the region $x \in (-1,0)$, as shown in Fig.~\ref{fig:plots}(b).
{These last two regions can be interpreted as below and above threshold for the production of two massive quarks, see e.g. \cite{Henn:2016tyf}.}
For the real part for $x \in (-1,0)$, the deviation can reach $25\%$ near $x=-0.2$, see Fig.~\ref{fig:plots}(d).
Still, we find it remarkable that the simple one-loop function $A(x)$ captures the main features of the four-loop result. 

\newpage

Let us now expand our novel results in several interesting limits. This serves both as a check and allows us to produce novel results, by computing additional terms. 
\begin{enumerate}[wide, labelwidth=!,  labelindent=0pt]
\item In the small angle limit $\phi \to 0$, i.e. $x \to 1$, 
we find agreement with known terms up to $\phi^6$ \cite{Grozin:2017css,Grozin:2018vdn}\footnote{The sign of $\alpha^4 n_l^{2}$ in eq. (4.2) of \cite{Grozin:2018vdn} should be reversed.}. The leading term is 
$B =\left( \frac{5 \pi^2}{54} + \frac{5 \pi^4}{108} - \frac{4 \pi^2 \zeta_{3}}{9} \right) \phi^2 +\mathcal{O}(\phi^4)$.
\item In the light-like limit $x \to 0$, we find
\begin{align}
B \; 
= &
-  \log x \left( \frac{\pi^2}{6} - \frac{\zeta_3}{3} - \frac{5 \zeta_5}{3} \right)    +\frac{5 \pi^2}{8} - \frac{11 \pi^4}{36} +\frac{53 \pi^6}{2835} \nonumber \\
& \hspace{-0.0 cm} - \frac{35 \zeta_3}{12} -\frac{ \pi^2 \zeta_3}{6}
- 3 \zeta_3^2 + \frac{185 \zeta_5}{12}  + \mathcal{O}(x)\,.
\end{align}
The first $\log x$ term agrees with the light-like cusp anomalous dimension \cite{Lee:2019zop,Henn:2019rmi}, and the finite part is new.
\item In the anti-parallel lines limit $x \to -1$, the cusp anomalous dimension is related to the quark-antiquark potential $V$ \cite{Kilian:1993nk},
$\Gamma_{\rm cusp}  \stackrel{\delta \to 0}{\longrightarrow} - C_{R} \frac{\alpha_{s}}{\delta} V$, where $\delta = \pi-\phi$. 
This relation is true up to beta function terms \cite{Grozin:2015kna}, which do not affect the quartic Casimir term considered here.
We find
\begin{align}\label{Bqqbarlimit}
B  \; = \; & -\frac{\pi}{\delta}  \Big( \frac{79  \pi^2}{72}  -  \frac{23 \pi^4}{48} 
+\frac{5 \pi^6}{192} +\frac{l_2 \pi^2}{2} + \frac{\l_2 \pi^4}{12}  \nonumber \\
&\quad\; - \frac{l_2^2 \pi^4}{4} -\frac{61 \pi^2 \zeta_3}{24} +\frac{21 \pi^2 \zeta_3 l_2}{4} 
\Big)   + {\mathcal{O}}(\delta) \,,
\end{align}
where $l_2 = \log(2)$. 
The pole term is in agreement with \cite{Lee:2016cgz}.
It is remarkable that the subleading terms in eq. (\ref{Bqqbarlimit}) start at ${\mathcal{O}}(\delta)$ only.
\end{enumerate}

\subsection{Quark-antiquark potential in $\mathcal{N}=4$ sYM}

The quark-antiquark potential in $\mathcal{N}=4$ sYM is known up to two loops 
for both the bosonic and supersymmetric Wilson loop \cite{Drukker:1999zq, *Erickson:1999qv, *Pineda:2007kz, *Drukker:2011za, *Prausa:2013qva}.
At three loops, it has the following color dependence,
\begin{align}
V_\textrm{sYM}|_{\alpha_s^3} = \left( \frac{\alpha_s}{\pi}\right)^3 \left[ C_{A}^3 V_{1} + {d_{R} d_{A}}/{(N_{R} C_{R})} V_{2} \right] \,.
\end{align}
We can determine $V_{2}$ for bosonic static charges by using a supersymmetric decomposition, as in \cite{Henn:2019swt}.
Taking into account the known results for the gluon and the fermion quartic Casimir terms \cite{Lee:2016cgz},
we only need the scalar contribution. We obtain the latter from our result $C(x)$ in eq. (\ref{definitionB}). (The full formula for $C(x)$ is provided in an ancillary file.)
We find
\begin{align}
V_{2}   =& \,  7 \pi^2 - \frac{47 \pi^4}{24} +  \frac{413 \pi^6}{1440} + 
 \frac{116   \pi^2 l_2 }{3} +  \frac{3 \pi^4 l_2}{3}  + \frac{2}{3} \pi^4  l_2^2 \nonumber \\ & \hspace{-0.5cm}- 
 \frac{17}{12} \pi^2  l_2^4 - 34 \pi^2 {\rm Li}_{4}\left(\frac{1}{2}\right)   - 
 \frac{89}{4}  \pi^2 \zeta_{3} - 14 \pi^2 l_2 \zeta_{3} \,.
\end{align}
This constitutes the first non-planar correction to
the quark-antiquark potential in $\mathcal{N}=4$ sYM.
It would be interesting if integrability methods \cite{Correa:2012hh, *Drukker:2012de, *Gromov:2016rrp} could be extended to this case.

\section{Discussion and conclusions}

We computed the matter-dependent quartic Casimir term of the four-loop angle-dependent cusp anomalous dimension, in a generic gauge theory involving massless matter fields. 
In particular, this determines the full QED result for this quantity, and includes for the first time contributions from light-by-light scattering diagrams.

Our calculation revealed new structures compared to previous three-loop results: the answer contains four as opposed to previously two rational structures.
The function space is given by iterated integrals, and compared to three loops, there is one new integration kernel, namely $d \log (1+x^2)$, in addition to $ d\log x$, $d\log(1-x)$ and $d\log(1+x)$.
This relative simplicity is remarkable given that intermediate steps (and higher order terms in the dimensional regulator) contain further integration kernels, cf. eq. (\ref{canonicalDE}). 
This hints at better approaches that avoid the complicated intermediate terms.

The information on the function space is valuable input for bootstrap approaches \cite{Almelid:2017qju}. In particular, our result implies constraints (via collinear limits) on the multi-leg soft anomalous dimension matrix for massive particles, which currently is known at two loops only \cite{Ferroglia:2009ep,Mitov:2009sv,Chien:2011wz}.

We analyzed the novel four-loop result numerically, and found that, surprisingly, it is described within a few per cent by a rescaled one-loop function (and up to $25 \%$ for the real part in the above threshold region). Clearly this could be improved even further by using some of the known limiting behaviour, as in \cite{Davies:2019roy}, for example. It would be interesting to understand why the approximation works so well here, in view of other problems where fully analytic results are not yet known, such as scattering processes with many mass scales.

To obtain all color contributions of the complete QCD cusp anomalous dimension at four loops, only two further contributions are needed \cite{Bruser:2019auj}. 
The first one is the planar limit, which is conceptually easier compared to our calculation. 
The HQET integrals we computed should cover a large part of the integrals needed, and are available upon request from the authors.
The second one is the gluonic quartic Casimir term. Thanks to our calculation of the scalar terms, the latter can equivalently be obtained from the non-planar $\mathcal{N}=4$ sYM contribution. This opens up novel approaches to this problem \cite{Henn:2019swt}.

\section{Acknowledgments}

This research received funding from the European Research Council (ERC) under the European Union's Horizon 2020 research and innovation programme (grant agreement No 725110), {\it Novel structures in scattering amplitudes}. R.B. was supported by the Deutsche Forschungsgemeinschaft (DFG) under grant  396021762 - TRR 257.

\newpage

\begin{widetext}

\appendix

\section{Integrals beyond multiple polylogarithms?}

Two of the integral families contain the integral sector shown in Fig.~\ref{fig:sector15}. It contains $11$ coupled basis integrals. Following the procedure outlined in the main text, we find the following obstacle to reaching a canonical form for the DE.
At order $\eps^0$ we encounter the following  two-by-two system
 (after eliminating spurious singularities \cite{Lee:2014ioa}),
\begin{align}
d  \left( \begin{array}{c}  g_1 \\ g_2 
\end{array}  \right) =  d  \left( 
 \begin{array}{cc} 
   -\frac{i}{2 \sqrt{3}}  \ln \frac{ y-y_{+} }{y-y_{-} }&      \frac{2}{3}   \ln y  +   \frac{1}{6}  \ln ( y-y_{+})  ( y-y_{-})        \\   
 \frac{1}{2}  \ln ( y-y_{+})  ( y-y_{-})    &  \frac{i}{2 \sqrt{3}}  \ln \frac{ y-y_{+} }{y-y_{-} }  \end{array} 
 \right)  \left( \begin{array}{c}  g_1 \\ g_2 
\end{array}  \right),
\end{align} 
where $y \equiv - \frac{(1-x)^2}{x}$ and $y_\pm =  3 \sqrt{3} \, e^{ \pm \frac{i \pi}{6}}$.
While the residue matrices at the singular points $0, y_{\pm}$ have eigenvalues $0$, at $y=\infty$ we find eigenvalues -1 and 1.
We were not able to find a balancing transformation \cite{Lee:2014ioa} that sets all eigenvalues to zero. 
It would be interesting to prove that this two-by-two system cannot be balanced via algebraic transformations, as it would mean that this integral sector involves functions beyond multiple polylogarithms.

\end{widetext}

 \bibliographystyle{apsrev4-1}

\bibliography{BibFile}

\end{document}